\documentstyle[aps,prb,multicol,ifthen,psfig]{revtex}

\newcommand{\wide}[2]{                                                                                             %
\end{multicols}                                                                                                                 %
\widetext                                                                                                                            %
\noindent                                                                                                                           %
\ifthenelse{\equal{#1}{t}}                                                                                              %
{}                                                                                                                                           %
{                                                                                                                                            %
\raisebox{0.1in}[0in][0.02in]{$\rule{3.575in}{0.002in}                                            %
\rule{0.002in}{0.08in}$}                                                                                                  %
}                                                                                                                                            %
#2                                                                                                                                         %
\ifthenelse{\equal{#1}{b}}                                                                                             %
{}                                                                                                                                           %
{                                                                                                                                            %
{\raisebox{-0.1in}[0in][0.02in]                                                                                       %
{\hspace{3.575in}$\rule{0.002in}{0.08in}                                                                   %
\rule[0.08in]{3.575in}{0.002in}$}                                                                                   %
}                                                                                                                                             %
}                                                                                                                                             %
\begin{multicols}{2}                                                                                                         %
\noindent                                                                                                                            %
}                                                                                                                                             %

\begin{document}

\title{Weak localization in InSb thin films heavily doped with lead}

\author{M.~Oszwa\l dowski\cite{email} and T.~Berus}
\address{Institute of Physics, Pozna\'n Technical University,
Nieszawska 13a, 61-022 Pozna\'n, Poland\\
and International Laboratory of High Magnetic Fields and Low Temperatures,
Gajowicka 95, 53-529 Wroclaw, Poland}
\author{V.~K.~Dugaev}
\address{Electronics and Communications Department, Instituto Superior de
Engenharia de Lisboa,\\
Rua Conselheiro Emidio Navarro, P-1949-014 Lisbon, Portugal\\
and Institute for Materials Science Problems, Ukrainian Academy of Sciences,
Vilde 5, 58001 Chernovtsy, Ukraine}
\date{\today }
\maketitle

\begin{abstract}
The paper reports on the investigations of the weak localization (WL)
effects in 3D (about 2~$\mu $m thick) polycrystalline thin films of InSb.
The films are closely compensated showing the electron concentration $%
n>10^{16}$~cm$^{-3}$ at the total concentration of the donor and acceptor
type structural defects $N_t>10^{18}$~cm$^{-3}$. Unless Pb-doped, the InSb
films do not show any measurable or show very small WL effect at 4.2~K. The
Pb-doping to the concentration of the order of $10^{18}$~cm$^{-3}$ leads to
pronounced WL effects below 7~K. In particular, a clearly manifested
spin-orbit (SO) scattering is observed. That is ascribed to the large mass
of the Pb atoms. From the comparison of the experimental data on temperature
dependence of the magnetoresistivity (MR) and sample resistance with the WL
theory, the temperature dependence of the phase destroying time $\tau
_{\varphi }$ is determined. The determination is performed by fitting
theoretical terms obtained from Kawabata's theory to experimental data on
magnetoresistance. It is concluded that the dephasing process is connected
to three separate interaction processes. The first is due to the SO
scatterings and is characterized by temperature-independent relaxation time $%
\tau _{so}\simeq 10^{-12}$~s. The second is associated with the
electron-phonon interaction ($\tau _i\sim T^{-3}$). The third dephasing
process is characterized by independent on temperature relaxation time $\tau
_c=(1$ to $7)\times 10^{-12}$~s. This relaxation time is tentatively
ascribed to inelastic scattering at extended structural defects, like
grain boundaries. The resulting time $\tau _{\varphi }$ shows saturation in
its temperature dependence for $T\rightarrow 0$. The temperature dependence
of the resistance of the InSb$<$Pb$>$ films can be explained by the
electron-electron interaction for $T<1$~K, and by the WL effect for $T>2$~K.\\
\\
PACS numbers{73.20.Fz; 72.15.Rn; 72.10.Fk.}
\end{abstract}

\begin{multicols}{2}

\section{Introduction}

Because of its unique physical properties InSb is one of the most
interesting materials for the investigation of the weak localization (WL)
effects. The critical electron concentration $n_{cr}$ for the
metal-insulator transition (MIT) in InSb is about $10^{14}$~cm$^{-3}$, which
is the smallest value among semiconductors.\cite{ootuka} This small critical
concentration results in a metallic conduction that can be realized with the
help of donor doping, in the electron concentration range that spans over
five orders of magnitude. This peculiar property allows the realization of
various physical situations under which the localization effects can be
investigated and compared with the existing theories. For the impurity
concentrations below $10^{19}$~cm$^{-3}$, the $\Gamma $ conduction band
structure is not disturbed meaningfully and hence in the interpretation of
the experimental data the known material parameters of structurally perfect
InSb can be exploited.

Among those parameters the most important for the effect in question is the
small effective mass at the bottom of conduction band, $m^{*}=0.014\, m_e$ ($%
m_e$ is the free electron mass) and high electron mobility which is very
sensitive to the presence of structural defects. The small electron
effective mass corresponds to a low density of states at the bottom of the
conduction band. In such case, even at relatively low electron density, the
condition of applicability of the WL theory, $E_F\tau /\hbar \gg 1$, is fulfilled
($E_F$ is the Fermi energy and $\tau $ is the elastic scattering time).

The small effective mass in conjunction with a large static dielectric
constant, $\varepsilon _s=17.5$, causes that the effective Bohr radius $%
a_B^{*}$ is very large, $a_B^{*}=\varepsilon _s\hbar ^2/m^{*}e^2\simeq 73$%
~nm. As a consequence, the non-ideality parameter, which determines a
relative importance of the interaction effect, $k_F\, a_B^{*}\simeq 4k_F^2\,
\lambda ^2$ is large for $n>n_{cr}$ ($\lambda $ is the screening length). In
such a case the corrections from interaction become smaller then those from
localization. We estimate that the transition from interaction to
localization occurs in the electron concentration range $10^{14}-10^{15}$~cm$%
^{-3}$.

In spite of those particular advantages as offers the metallic InSb, only
several investigation teams studied the WL effects in this semiconductor.
\cite{dynes,morita,mani,aronzon} In all papers, bulk crystals of InSb with
electron concentrations $n<5\times 10^{15}$ cm$^{-3}$ were investigated. The
studies revealed a divergence in the interpretation if the corrections are
related to localization or electron-electron interactions. The WL effects
were observed at temperatures smaller than 1~K and their strength was
smaller than that predicted by the theory.

In the present paper we investigate the WL effects in three-dimensional (3D)
polycrystalline thin films of InSb heavily doped with electrically neutral
lead. The background effective donor concentration in these films is between
$1.6\times 10^{16}$~cm$^{-3}$ and $2\times 10^{17}$~cm$^{-3}$, which means
that we investigate samples with much higher concentration than in previous
studies. Hence, we expect that for the present concentrations the
corrections should be related mostly to localization effects. Note that the
results of Ref.~[\onlinecite{morita}] have demonstrated that the corrections
are due to the interaction.

In our samples of InSb$<$Pb$>$ films, the WL effect is observed at relatively
high temperatures, $T\leq 7$~K and is also much more pronounced in the
magnetoresistance (MR) as compared to the earlier observations. A
particular feature of the samples is a clearly marked effect of the SO
interaction (positive MR) at the lowest magnetic fields. The SO scattering
was not observed in the previous investigations performed on InSb doped with
shallow donors. The presently observed SO interaction seems to be associated
with the doping with Pb, and in such a case is due to its large atomic
number. The enhancement of the WL in InSb$<$Pb$>$ films can be ascribed to
the scattering of electrons from short-range potentials of Pb atoms
incorporated substitutionally into the InSb lattice in pairs for neighboring
In-Sb pairs (we discuss it in Sec.~III).

For the interpretation of our experimental results we had to complete the
commonly accepted formulae for the WL corrections\cite{alt82,lee85,alt85} to
take into account the specific features of InSb. They are a strong SO
interaction and a strong Zeeman splitting of spin up and down electrons in
magnetic field (Land\'e factor for InSb is $g=-51.3$). These effects have
been discussed separately in Refs.~[\onlinecite{alt81,lee82,maekawa,isawa82}].
In particular, in a theoretical work of Isawa et al.,\cite{isawa82}
dedicated to the localization and interaction effects in InSb, only Zeeman
splitting has been taken into account, whilst SO coupling has not. Since our
MR experiments show the importance of SO coupling, we have to make necessary
generalizations of the known formulas to take into account both SO
scattering and Zeeman splitting.

The investigation of the WL effects have acquired a new impact in recent
years. This is, to a great extent, due to the problem of the temperature
dependence of the electron dephasing time (for a quick introduction of the
problem and relevant literature consult Ref.~[\onlinecite{zawadowski99}]). Our
paper is also relevant to this problem.

The paper is arranged as follows.
In Section~II we describe the preparation method of InSb films and their
doping with Pb.
Section~III is devoted to the interpretation of the inherent electrical
properties of the InSb films, which are associated with the presence of
electrically charged structural defects and their non-uniform spatial
distribution.
In Section~IV we discuss the WL effects observed below 7~K.
Section~V summarizes our finding and conclusions.

\section{Sample preparation and characterization}

Thin films of InSb for the present investigations were obtained by the flash
evaporation method. High purity (zone refined) InSb with the net donor
concentration $N_d-N_a<5\times 10^{15}$~cm$^{-3}$ was used for the
evaporation. The films were evaporated on ceramic substrates through
mechanical masks and were shaped for conductivity and Hall effect
measurements. They have thickness of about 2~$\mu $m. Details of these
experimental procedures are given elsewhere.\cite{osz84}

The obtained polycrystalline InSb films have rather typical properties for
the given preparation conditions. Such films have electrical properties
governed by their structural defects.\cite{wieder} Without any intentional
doping the films are of $n$-type. With the increase in substrate temperature
the number of structural defects decreases and the electron concentration
decreases, whilst the electron mobility increases. The mobility is also
dependent on film thickness: below 3~$\mu $m it markedly decreases with the
film thickness. In the thicker films obtained at substrate temperatures $%
T_s\geq 450^{\circ }$C, the lowest electron concentration is of about $%
2\times 10^{16}$~cm$^{-3}$, and the highest room temperature electron
mobility is in excess to 10000~cm$^2$/V$\cdot $s. For this electron
concentration the mobility is a half-of-order magnitude smaller than that in
uncompensated bulk crystals of InSb. The large number of structural defects
of the donor and acceptor type, which partly compensate each other, can
explain this mobility drop. Below room temperature the mobility decreases
with temperature, which can be explained by the barrier mechanism (see
Sec.~III).

Lead doping was performed by the co-evaporation of InSb and Pb from separate
evaporation sources. This procedure was essentially the same as that used
for doping of InSb films with tin.\cite{ob} The lead doping, up to the
solubility limit, which is of the order of 0.02 atomic percents ($3\times
10^{18}$~cm$^{-3}$), has no electrical influence on the doped InSb films,
except for the lowest temperatures. We explained this electrical neutrality
of lead by a particular incorporation into the InSb lattice.\cite{ob} Lead
is an amphoteric impurity in InSb and incorporates to the InSb matrix in
pairs substituting the neighbor pairs In$-$Sb. In this sense Pb is an
isoelectronic impurity in InSb. However, in our samples the solubility limit
is overrun and in result the InSb$<$Pb$>$ films have microscopic inclusions
of lead. Detailed investigations suggest\cite{ob,obd} that a part of those
inclusions can transit to superconductivity below 5~K.

We investigate four samples of InSb$<$Pb$>$: samples 732, 725A and 725B,
obtained at $T_{s}=370^{\circ }$C, and sample 727, obtained at $%
T_{s}=340^{\circ }$C. Their basic electrical parameters, as obtained by the
Hall measurements at 4.2~K, are given in Table 1. Samples 725A and 725B were
obtained at the same evaporation run and their electrical parameters are
very similar. For this reason in Table 1 we present only the parameters of
sample 725B.

In Table 1, $d$ is the thickness of InSb film, $\mu $ is the electron
mobility and $n$ is the electron concentration. The meaning of other
parameters is explained in Sec.~III.

The dependence of resistivity of all InSb$%
<$Pb$>$ films on temperature and magnetic field is of the same character,
but is different in details. In Fig.~1 we present results for samples 725B
and 727 below room temperature. These temperature dependences are typical
for polycrystalline InSb thin films, except the low temperature region below
7~K. The sharp resistivity increase with the temperature decrease in this
region, shown in the inserts, was observed only in Pb-doped InSb films. We
ascribe this low-temperature increase in resistivity to the WL effect (see
Sec.~IV,B). It is worth noting that the transition to the localization does
not always have such an abrupt form as those seen in the inserts. For
example, this transition for sample 725A has a smoother form.

In the considered temperature range, the electron density in the InSb films
is independent of temperature, if we neglect a small decrease in the
vicinity of room temperature. Thus, the increase in resistivity is solely
due to a decrease in the mobility. This decrease in mobility as well as the
stronger temperature dependence of the mobility in samples with smaller
electron concentrations is rather typical for polycrystalline InSb thin
films.\cite{osz84,wieder}

The transversal and longitudinal magnetoresistivity of the samples at 4.2~K
is presented in Fig.~2. The observed MR, except for the low field region
shown in the insert, is also typical for InSb films. The negative MR in the
low field region is characteristic for Pb-doped InSb films. Actually, in a
narrow temperature range, and in magnetic fields below 0.1~T, the MR is
positive (Fig.~3), but
in the course of measurements shown in Fig.~2, this region has not been
traced and for this reason the positive MR does not appear in the figure.

The sample investigations in the temperature range 1.7 to 295~K have been
performed at the International Laboratory of High Magnetic Fields and Low
Temperatures in Wroclaw, and the investigations in the temperature range
0.04 to 2~K have been carried out at the Institute of Low Temperatures and
Structural Investigations in Wroclaw, Poland.

\section{Interpretation of electrical properties of
I\lowercase{n}S\lowercase{b} films and effect of P\lowercase{b} doping}

As mentioned above, the dependence of sample resistance on temperature or
magnetic field, except for the lowest temperature and fields, is
substantially the same in InSb films doped and undoped with Pb. The low
temperature effects below 7~K, recognized as the WL, will be discussed in
Sec.~IV. First we discuss the properties, which are common for InSb and InSb$%
< $Pb$>$ films, and then we consider the role of Pb doping.

The temperature dependence of the resistivity of polycrystalline InSb thin
films down to 7~K can be explained by a large concentration of electrically
active structural defects that scatter electrons. These defects are
impurities and other point defects, such as interstitials, vacancies,
stacking faults, etc. If a possible non-uniformity in the distribution of the
ionized defects is ignored, the low temperature mobility can be described by
the Brooks-Herring formula that takes into account the conduction band
non-parabolicity of InSb.\cite{kolodziejczak} For an electron concentration $%
n$, the formula gives the following mobility in a compensated sample:
\begin{equation}
\label{1}
\mu _i=\mu _{io}(n)\, \frac n{n+2N_a}\; ,
\end{equation}
where $\mu _{io}(n)$ is the electron mobility limited by the ionized
impurities in an uncompensated sample, and $N_a$ is the concentration of
acceptors. For $\mu _{io}(n)$ we use the experimental data, instead of the
theoretical ones, to avoid the question about the validity of the classical
theory at very low temperatures.

Consider InSb$<$Pb$>$ thin film with a typical electron concentration of $%
2\times 10^{17}$~cm$^{-3}$ and mobility $\mu =1000$~cm$^2$/V$\cdot $s. For
this concentration, a low temperature $\mu _{i0}(n)$ is $(4$~to~$5)\times
10^4$~cm$^2$/V$\cdot $s.\cite{gershenzon} Inserting this value into Eq.~(1),
one obtains the acceptor concentration $N_a\geq 4\times 10^{18}$~cm$^{-3}$.
This value seems to be too high, as the resulting high degree of
compensation $K=0.95$ is rather improbable for an accidental compensation
process. Hence, we assume that the small mobility is only partly due to the
compensation effect.

Since InSb has an isotropic conduction band, any classic longitudinal MR
should not be observed. Also the magnitude of the positive transversal MR is
higher than that expected for uniformly doped InSb having the same electron
concentration. The observed longitudinal MR as well as the relatively large
transversal MR have to be associated with a non-uniform distribution of the
charged defects. These effects of non-uniform distribution have been
predicted by Herring.\cite{herring} However, assuming reasonable
fluctuations of the electron density, the theory predicts much smaller
longitudinal MR than we observe. In order to explain the observed effects,
we assume that the polycrystalline InSb thin films show not only a large
number of defects of the donor and acceptor type but also that their
compensation degree fluctuates strongly. Such fluctuations can for example
be generated at grain boundaries.

With this assumption we can adopt to our samples the explanation of the
galvanomagnetic properties of compensated bulk $n$-InSb proposed by Aronzon
et al.\cite{aronzon} In $n$-InSb, due to the low density of states and high
dielectric constant, the compensation fluctuations can create macroscopic
potential fluctuations. In result, the electrons are located in potential
wells separated by some potential barriers. Thus, the electric conduction
takes place in a material composed of macroscopic areas of a metal-like
phase separated by regions of an insulating phase. The conduction is of
percolating character and can be described by a percolation level $E_p$.
At $T=0$ the conduction can take place only when the Fermi level $E_F>E_p$.

In this model, the measured macroscopic mobility is thermally activated and
hence the resistivity decreases with increasing temperature. At low
temperatures, the microscopic mobility (we define the microscopic mobility
as that inside the potential wells) also decreases with temperature as it is
mostly limited by the scattering from charged defects. However, the
macroscopic mobility, measured in the Hall experiments, is mainly limited by
the percolation. In such a case the current lines are of zigzag shape, and
in result the measured MR has both the transversal and longitudinal
component. Therefore, the assumed percolative conduction model explains both
the temperature dependence and magnetic field dependence of the sample
resistivity, shown in Figs.~1 and 2, respectively.

For small values of $E_F-E_p$, i.e. closer to the MIT of the percolative
type, the temperature and field dependence of the resistivity should be
stronger. Closer to the MIT are the samples with smaller electron
concentration. As is seen in Figs.~1 and 2, sample 725B having smaller $n$
shows stronger dependencies of $\rho (T)$ and $\rho (B)$ indeed.

We now examine the low-temperature region where the WL effects occur. Since
pronounced localization effects take place only in samples doped with Pb, we
also have to consider a possible effect of Pb-doping on both elastic $\tau $
and inelastic $\tau _{\varphi }$ relaxation times (here $\tau $ is defined
as the microscopic magnitude). When $\tau $ changes due to the doping with
Pb, the macroscopic mobility does not change substantially since it is
mainly limited by the percolation. Therefore, in frame of this model, we do
not rule out a decrease in the microscopic mobility due to Pb doping.
However, this decrease can not cause any dramatic changes in the WL
observation, because the WL correction in 3D case depends on the relaxation
time as $\sqrt{\tau}$ . \cite{alt82,lee85} We believe that to understand the
role of Pb doping, it is important to take into account that the neutral Pb
atoms interact with electrons by short-range potentials, like strongly
screened ionized impurities in metals. Large number of such defects creates
necessary conditions for diffusive motion of electrons, which results in the
WL effect. The suggestion that the WL is related to the electron scatterings
on Pb atoms can further be confirmed by the discussed in Sec.~IV
antilocalization effect due to the SO interaction. According to the theory,
\cite{abrikosov62} the SO scattering depends on the impurity atom mass as
$Z^{4}$. More recent theoretical results for Mn suggest that the dependence
be as $Z^{2}$ and the scattering can also depend on the impurity valency.
\cite{papanicolaou} Thus, both the theories predict strong SO scattering
from heavy impurities. It should be mentioned that in previous studies
performed on InSb crystals\cite{dynes,morita,mani,aronzon} no SO effects have
been observed.

In principle, there is a possibility that the Pb doping increases the
inelastic relaxation time associated with the electron-phonon interaction,
by changing the phonon density of states. It would increase the localization
corrections.\cite{alt82,lee85} However, this possibility should be ruled out
because the corresponding changes in the low-energy phonon spectrum turn out
to be too small to change meaningfully the relaxation time $\tau _\varphi $.
Besides, the Pb doping should rather decrease than increase the inelastic
relaxation time because heavy impurities usually increase the phonon density
of states on the low energy side.\cite{kagan}

\section{Weak localization effects}

The important parameters of our samples from the viewpoint of the WL are
given in columns 5-9 of Table 1. The values of parameters estimated from the
Hall measurements at $T\simeq 7$~K are presented without parenthesises. There
$v_F$ is the electron Fermi velocity, $k_F$ is the Fermi wavevector, and $l$
is the mean free path of electrons. The physical meaning of other parameters
is the following. $1/k_Fl$ is a basic parameter of the WL theory. The WL
theory works for $(1/k_Fl)\ll 1$. $r_s$ is the interaction parameter.
Interaction is expected to be less important if $r_s\ll 1$. $B_c$ is a
critical field in Kawabata's theory of magnetoconductivity.\cite{kawabata}
For $B>B_c$ the localization corrections are suppressed completely by the
magnetic field. $l_{\varphi }=\left( D\tau _{\varphi }\right) ^{1/2}$ is the
dephasing mean free path. $D=v_Fl/3$ is the diffusion constant in 3D case.
For the estimation of $l_{\varphi }$ we assume $\tau _\varphi =10^{-12}$~s
at 4.2~K. The detailed analysis (see Sec.~III,A) confirms this value.

From Table 1 one can conclude the following. The condition of effective
three-dimensionality,\cite{alt81,lee85} $l_{\varphi }\ll d$, is fulfilled in
all our films. However, for sample 732 the condition $1/k_{F}l\ll 1$ is not
fulfilled. This means that for that sample the WL theory might not be
applicable. Also, for this sample the magnetic freezing critical field (see
below) turns out to be lower than for the other samples. However, even in
this case the magnetic freezeout is not likely to occur.

In the magnetic field region of interest, $B<0.3$~T, the classical
magnetoconductivity of the samples is expected to be negligible. This is
because in this magnetic field region the condition of classically weak
magnetic fields, $\mu B\ll 1$, is fulfilled for all samples.

In all studied samples the condition $l\ll a_B$ is fulfilled. This condition
is important from the point of view of applicability of the interaction
theory in its usual form.\cite{alt82,alt85} Our estimations show that the
inequality $l\ll a_B$ can change the temperature dependence of resistivity
at $\hbar/\tau <kT<E_Fl/a_B$.

In the case of InSb a question of the critical field for the magnetic
freezeout of electrons at impurities may be risen. This magnetic field can
be estimated from the relation $n(a_B\, l_B^2)^{1/3}\simeq 0.34,$\cite
{morita} where $l_B=(\hbar/eB)^{1/2}$ is the magnetic length. In our
samples, such a critical field is higher than 2~T, and thus the freezeout
does not take place at the fields in question.

\subsection{Magnetoresistance}

In all samples the dependence of resistivity on magnetic field for $B<0.3$~T
and $T<7$~K, is such that it first increases and then decreases with
magnetic field. A typical example of this behaviour is shown in Fig.~3. The
shown experimental data on magnetoresistance are related to the WL effect
exclusively. A small contribution to the magnetoresistance from the
classical (Lorentz) effect is subtracted on the basis of the extrapolation
of the classical magnetoresistance from the higher magnetic fields (Fig.~2).

The clearly seen positive MR region at the lowest magnetic fields can be
explained within the WL theory as the result of strong SO scattering.\cite
{bergmann,alt81} The negative MR, seen at magnetic fields higher than 0.05~T
is the normal behavior expected for the case of weak localization when the
SO interaction (the antilocalization) can be neglected.

In order to perform theoretical curve fitting to the experimental points,
with the aim of determination of the sample parameters, we adopt the usual
Kawabata's approach to the magnetoresistivity in 3D case.\cite{kawabata} The
original formulae for the conductivity corrections have to be generalized to
take into account both SO coupling and Zeeman splitting. We find that the
contribution from singlet Cooperon that contains Zeeman splitting is
\wide{m}{
\begin{eqnarray}
\label{2}
\Delta \sigma _{s}(B)-\Delta \sigma _{s}(0)
=\frac{e^2}{4\pi ^2\hbar l_B}\sum\limits_{n=0}^{\infty }
\left\{
\frac{\cos \left( \varphi _1/2\right) }
{\left[ \left( n+1/2+\delta \right) ^2+\nu ^2\right] ^{1/4}}
\right. \nonumber \\
\left.
-2\cos \left( \varphi _2/2\right)
\left[ \left( n+1+\delta \right) ^{2}+\nu ^{2}\right] ^{1/4}
+2\cos \left( \varphi _0/2\right)
\left[ \left( n+\delta \right) ^{2}+\nu ^{2}\right] ^{1/4}
\right\}
\end{eqnarray}
and that from triplet Cooperon, which contains SO scattering is
\begin{equation}
\label{3}
\Delta \sigma _{t}(B)-\Delta \sigma _{t}(0)=-\frac{3e^{2}}{4\pi ^{2}\hbar
l_{B}}\sum\limits_{n=0}^{\infty }\left[ \frac{1}{\sqrt{n+1/2+\delta _{so}}}
-2\left( \sqrt{n+1+\delta }_{so}-\sqrt{n+\delta _{so}}\right) \right] .
\end{equation}
}
Here we denoted: $\delta =l_{B}^{2}/4D\tau _{\varphi }$ , $\delta
_{so}=\left( l_{B}^{2}/4D\right) \left( 1/\tau _{\varphi }+4/3\tau
_{so}\right) $ , $\nu =g\mu _{B}/4eD$,
$\varphi _0=\arctan \left[ \nu /\left( n+\delta \right) \right] $,
$\varphi _1=\arctan \left[ \nu /\left( n+1/2+\delta \right) \right] $,
$\varphi _2=\arctan \left[ \nu /\left( n+1+\delta \right) \right] $,
and $\mu _B$ is the Bohr magneton.
\newline
For shortness we imply here $\tau _{\varphi }$ is the total phase relaxation
time. The total localization correction to conductivity is the sum of (2)
and (3).

We used Eqs.~(2) and (3) for the fitting of theoretical curves to the
experimental dependence of the magnetoresistivity $\rho (B)/\rho (0)$ at low
magnetic fields and various temperatures. In the calculations we take the
measured conductivity $\sigma =\sigma _0+\Delta \sigma $, where $\sigma
_0=en\mu $ is the classical conductivity, which we assumed to be independent
upon magnetic field in the field region in question.

To fit the magnetoresistivity curves we had to choose four parameters
$\tau _{\varphi }$, $\tau _{so}$, $n$ and $\mu $ and their temperature
dependencies. The last two parameters determine the Fermi velocity and the
elastic relaxation time. We have assumed that $n$ is temperature independent
and $\mu $ is weakly dependent on temperature in the temperature range in
question. So, the observed weak temperature dependece of the samples
resistance is solely due to the weak temperature dependence of the mobility.
We have also assumed that $\tau _{so}$ is independent of temperature. This
results from the assumed temperature independence of the SO interaction and
the elastic electron free path. Such an assumption is frequently adopted.
\cite{wang}

An example of the theoretical curve fitting to experimental points obtained
for sample 727 is shown in Fig.~3. As is seen, a good agreement between
experiment and theory is reached only at the lowest magnetic fields. This is
acceptable, as Eqs.~(2) and (3) hold only for the lowest magnetic fields.
\cite{isawa82}

Sample parameters, obtained from the curve fittings, are given in Table~1 in
parentheses. They can be compared to those obtained from the Hall
measurements. In our attempts to fit the theoretical curves to the
experimental data we found it impossible to reach the values of
concentration and mobility obtained from the Hall measurements. In the case
of sample 732 we found an increase in the mobility by a factor of about 3.
This means that the microscopic mobility in this sample, as expected, is
considerably higher than the macroscopic one. With this higher microscopic
mobility, the condition $1/k_Fl<1$ is fulfilled, which confirms the
applicability of the WL theory to the description of the magnetoresistivity.

In the case of samples 725 and 727 we found a relatively small increase in
the mobility along with a considerable decrease in the electron
concentration. The origin of this disrepancy is not quite clear. We can
suppose that the Hall measurements were influenced by the Pb inclusions. Due
to a possible short-circuiting effects on the Hall voltage, the determined
electron concentration can be higher than the microscopic one.

In our curve fitting to the magnetoresistance measured at different
temperatures, we have assumed that the theoretical description has to use
the same set of material parameters $\tau _{so}$, $n$, and $\mu $. This
assumption appeared to be very restrictive and resulted in a considerable
uncertainity in the determination of $\tau _{\varphi }$ (see Fig.~4).
Moreover, to obtain improved fits we were forced to introduce into $\Delta
\sigma $ a temperature-dependent prefactor $\alpha (T)$. The value of
$\alpha $ at $T=4.2$~K is 1.0 for sample 732, 0.90 for sample 727, and 0.94
for sample 725B. The value of $\alpha (T)$ at $T=0.68$~K is 0.87 for sample
732 and 0.76 for sample 727. Sample 725B has not been measured at 0.68~K,
but at $T=1.8$~K it has $\alpha =0.81$. Therefore, the prefactor introduces
small corrections to the calculated curves mainly at the lowest temperatures.

It should be mentioned that in previous papers devoted to the WL effect in
InSb\cite{dynes,morita,aronzon} a prefactor of a constant value was also
introduced because the magnitude of the experimental magnetoresistance was
always smaller than that predicted by the theory. The value of the prefactor
was in the range $\alpha =0.02-0.55$. In the present case there is no
problem with the magnitude of the effect, but with its temperature
dependence. The temperature dependence of $\alpha (T)$ seems to be
correlated with the dose of Pb obtained by a given sample. The doses
obtained by samples 725 and 727 were higher than that obtained by sample
732. We think that the temperature dependence of $\alpha $ can be related to
this part of Pb inclusions that can transit to superconductivity. The effect
of the superconducting inclusions on the sample magnetoresistance was
disccussed in Ref.~[\onlinecite{ob}].

The high concentration of Pb in the InSb films may suggest that the
contribution of the superconducting fluctuations, known as Maki-Thompson
corrections,\cite{larkin} can play a significant role in the temperature
dependence of the conductivity. The Maki-Thompson corrections change $\Delta
\sigma $ by a factor $\beta (T)$ such that $0\leq \beta (T)\leq 1$. This
factor is related to $\alpha (T)$ as $\alpha (T)=1-\beta (T)$. However, the
temperature dependence of $\alpha $ in sample 727 can hardly be explained by
Maki-Thompson corrections because the corrections are rather smaller and
have different temperature dependence.

The temperature dependence of $\tau _{\varphi }$ and $\tau _{so}$ for the
four samples obtained from the fittings are presented in Fig.~4. The bars
attached to the values of $\tau _\varphi$ show the ranges where the curve
fitting is still reasonable, with the other parameters kept constant.

As is seen, the obtained SO relaxation time is nearly the same in all
samples and it is about $\tau _{so}=1.2\times 10^{-12}$~s. We can not
compare this value with any earlier data, because, as far as we know, these
are the first data on the SO scattering from impurities in conduction band
of a 3D semiconductor sample. Thus, we compare the scattering time with the
data for 2D metallic films covered with about a monolayer of impurities
compiled in Ref.~[\onlinecite{kobayashi}]. For the heaviest impurity atoms a
value of $\tau _{so}\leq 10^{-12}$~s is characteristic, which is close to our value.
It is believed that the probability of SO scattering is proportional to the
atomic number of the impurity and the inverse of the elastic relaxation
time. However, a detailed value is determined by the relevant conduction
band parameters.\cite{kobayashi} Thus, the relaxation times $\tau _{so}$ in
our samples should be proportional to the electron mobility. Actually, such
correlation in our samples is not observed. This can be explained by a
difference in the Pb concentrations between different samples. Such a
dependence of $\tau _{so}$ on the heavy impurity concentration was observed
in Mg films doped with Bi.\cite{wang}

As is seen in Fig.~4, at 4.2~K, $\tau _{\varphi }$ for all samples is of
the order of $10^{-12}$~s, and thus is close to the value of $\tau _{so}$.
However, the most striking feature of $\tau _{\varphi }$ is its weak
temperature dependence, when compared with earlier investigation results. It
should be pointed out that usually a dependence of the form
\begin{equation}
\label{4}
\tau _{\varphi }^{-1}\sim T^{p}
\end{equation}
is assumed for the inelastic processes. In particular, $p=2$ for the
electron-electron interaction and $p=3$ for the electron-phonon interaction
in the clean limit.\cite{alt81,rammer,reizer}

Dynes et al.\cite{dynes} found that in their samples $\tau _{\varphi }\sim
T^{-3}$ in the temperature range 0.05 to 1.5~K. Extrapolating this
dependence in Fig.~3 of Ref.~[\onlinecite{dynes}] to 4.2~K, one obtains $%
\tau _{\varphi }=1.3\times 10^{-12}$~s, a value close to our result. On the
other hand, Mani et al.\cite{mani} obtained for the temperature range 0.5 to
4.2 K, $\tau _{\varphi }\sim T^{-2}$, with $\tau _\varphi =4\times 10^{-12}$%
~s at 4.2~K.

It should be mentioned that the saturation of $\tau _{\varphi}$ at low
temperatures has been observed earlier many times at lowest temperatures,
and it had been usually attributed to the effect of magnetic impurities.\cite
{bergmann,larsen} This mechanism can not, however, be responsible for the
presently observed saturation of $\tau _{\varphi }$. This is because in our
samples important could be only magnetic impurities in concentration $%
>10^{17}$~cm$^{-3}$. Such a large uncontrolled doping should be ruled out from
the point of view of our technology of the sample preparation. Moreover, the
solubility limits of the typical magnetic impurities Fe, Co, Ni in InSb is
below 10$^{14}$~cm$^{-3}$.~~\cite{popov}

The saturation of $\tau _{\varphi }$ in InSb$<$Pb$>$ films has been earlier
demonstrated in Ref.~[\onlinecite{osz99}]. Meantime we performed further
experimental studies of \ the MR effect and we gathered further evidence for
the saturation. In these experiments we have paid a particular attention to
the possible thermal effect connected to the sample driving current. We are
convinced that the presented here saturation effect is not a result of the
current heating.

The saturation of $\tau _{\varphi }$ at low temperatures and in the absence
of magnetic impurities has been recently reported \cite{mohanty,lin} and
critically discussed in theoretical papers.\cite{alt98,golubev01,buttiker01}
While the existence of the saturation is still a \ subject of controversy,
it seems to emerge that an explanation of the observed weak temperature
dependence of $\tau _{\varphi }$ in some disordered metals is an additional
independent on temperature relaxation mechanism. A simple example of such a
mechanism is the inelastic scattering from two-level centers.\cite
{zawadowski99,fukuyama99} In our case we can assume that this mechanism is
associated with inelastic interactions of the conducting electrons in the
course of their transmission through potential barriers located at extended
structural defects like grain boundaries. We can expect the presence of
dangling bonds, acting as two-level centers, in the vicinity of the grain
boundaries. In this case, we should make the following substitution in the
relevant equations
\begin{equation}
\label{5}
\tau _{\varphi }^{-1}=\tau _{i}^{-1}+\tau _{c}^{-1}\;,
\end{equation}
where $\tau _{i}$ is the inelastic relaxation time related to intrinsic
mechanisms of relaxation that characterises uncompensated bulk InSb and $%
\tau _{c}$ is an independent on temperature relaxation time, which can be
related to the potential well of radius $L_{c}$, $\tau _{c}=L_{c}^{2}/D$.

Assuming that $\tau _i$ is of the form of Eq.~(4) with $p$ being an integer,
the best fit of the relation (5) to the points in Fig.~4 is obtained for $%
p=3 $. The solid curves in the figure are calculated from Eq.~(5) with
\begin{equation}
\label{6}
\tau _i=1.95\, \left( \frac{4.2}{T}\right) ^3 10^{-12}\; [s].
\end{equation}
The values of $\tau _c$ used in the calculations can be read out from Fig. 4
as the values of $\tau _{\varphi }$ at $T=0$. The dependence of the type of
Eq.~(6) represents the electron-phonon relaxation mechanism in pure metals.
The fact that the same dependence of Eq.~(6) can be used for the
calculations of all the curves is a further confirmation of relaxation time
mechanism, because the electron-phonon interaction is independent on the
electrical properties of the sample.

In pure metals, for $kT\gg \hbar c_l/l$, the electron-phonon relaxation time
can be calculated using the formula\cite{rammer,reizer}
\begin{equation}
\label{7}
\frac{1}{\tau _{e-ph}}
=\frac{7\pi }{10}\, \frac{(k_BT)^3}{\hbar mMc_l^4}\; ,
\end{equation}
where $m$ and $M$ respectively are the electron and ionic masses, and $c_l$
is the longitudinal phonon velocity. Assuming for InSb $c_l=3.8\times 10^3$
cm/s to be an averaged over crystallographic directions sound velocity,\cite
{landolt} we find that for the samples investigated the condition $kT\gg
\hbar c_l/l$ is fulfilled for $T>2.4$ K. If we assume that in the case of
semiconductors the electron mass in Eq.~(7) is the effective mass $m^*$,
then for InSb at 4.2~K we obtain $\tau _{e-ph}=5\times 10^{-11}$~s. This
value is by a factor of 25 higher than that obtained from our experiments.
Such discrepancy is however expected taking into account the fine
crystalline character of our samples. First, we expect that $c_l$ in our
samples is smaller than in InSb single crystal. Secondly, Eq.~(7) assumes
that the longitudinal and transversal phonon velocities are approximately
equal, $c_l\simeq c_t$ (in cubic crystals $c_l\simeq \sqrt{3}c_t$), and in
such a case the transversal phonons can be neglected. In strongly defected
samples $c_t$ can be considerably more reduced than $c_l$. In such a case
the transversal phonons can be important, which means that that the $e-ph$
relaxation time is further reduced.

\subsection{Temperature dependence of resistance}

The temperature dependence of the samples resistance between 0.04~K and 7~K
is displayed in Fig.~5. These dependencies are interpreted using the WL
theory, taking into account both localization and interaction corrections.

The localization correction to the conductivity at three dimensions ($d=3$),
which includes SO coupling, can be obtained by a generalization of the
formula of Ref.~\onlinecite{alt82} \ to the form:
\begin{equation}
\label{8}
\delta \sigma _{loc}={\rm const}+\frac{e^{2}}{4\pi ^{2}\hbar \sqrt{D}}\left[
3\left( \frac{1}{\tau _{\varphi }}+\frac{4}{3\tau _{so}}\right)
^{1/2}-\left( \frac{1}{\tau _{\varphi }}\right) ^{1/2}\right] \; .
\end{equation}
The interaction correction is
\begin{equation}
\label{9}
\delta \sigma _{int}={\rm const}_{1}
+2.5\, \frac{\sqrt{2}\, e^{2}\, T^{1/2}}
{6\pi ^2\, \hbar \, D^{1/2}}\; .
\end{equation}
Here we take into account only the exchange diagram contribution,\cite{alt82}
and do not consider the Maki-Thompson corrections. It makes sense for the
weak electron-electron interaction, when $k_{F}\lambda \gg 1$.

In general, the resistance below 1~K exhibits a dependence of the type $%
R\sim T^{-1/2}$. This dependence is shown by dashed lines in Fig.~5.
Therefore, the electron-electron interaction is responsible for the
temperature dependence in this temperature region. The exception from the
rule is sample 727 that shows a maximum in its temperature dependence. We
ascribe this anomaly to the superconducting Pb inclusions, which can
decrease the resistance.

At higher temperatures, the temperature dependence of the samples resistance
can be described by Eq.~(8), in which the temperature dependence of $\tau
_{\varphi }$ is given by the solid curves in Fig.~4. The dependence $R(T)$
is shown in Fig.~5 by the solid lines. As is seen, between 2~K and 5~K the
character of the temperature dependence of the samples resistance is well
approximated by the calculated curves. Above 6~K the quantum corrections to
the resistance can be neglected. We draw attention to the fact that both the
magnetoresistivity at various temperatures and the temperature dependence of
the samples resistance can be explained with the help of the same
temperature dependence of $\tau _{\varphi }$ given by Eq.~(5), in which an
important role plays the temperature independent relaxation time $\tau _{c}$.

Sometimes one argues that the saturation of $\tau _{\varphi }$ is an
artifact associated with the warming effect of the sample driving current
(see the discussion in Ref.~[\onlinecite{lin}]). In this connection we wish to point
out the fact that in Fig.~5 the resistance sharply increases with decreasing
temperature in the low temperature region where the dephasing time
saturates. This confirms that the hot-electron effects due to Joule heating
do not take place. If Joule heating occurred, the temperature dependence of
the resistance would saturate.

\section{Summary and conclusions}

This paper reports on the first observation and analysis of the weak
localization effects in 3D InSb thin films heavily doped with lead. The
investigated InSb$<$Pb$>$ films are prepared by vacuum coevaporation of high
purity In, Sb and Pb from separate evaporation sources. They are of
polycrystalline nature and show the electron concentration
$n>10^{16}$~cm$^{-3}$, which is independent on temperature
in the low temperature region
(the metallic behavior). This concentration is equal to the net donor
concentration, and it is due to the structural defects. The total number of
donor and acceptor type structural defects is estimated for
$>10^{18}$~cm$^{-3}$. Thus, the presently investigated samples of InSb are closely
compensated and have both the electron and total defect concentrations
considerably higher than those in the reported previous investigations.

The concentration of lead atoms incorporated into the crystal lattice is
estimated to be also of the order of $10^{18}$~cm$^{-3}$. This concentration
corresponds to the solubility limit of Pb in InSb. Lead incorporates into
the lattice of InSb mainly in pairs substituting for neighboring In-Sb
pairs. In this sense it is an isoelectronic impurity that is electrically
neutral (it does not supply mobile carriers). The predominant lattice
incorporation in pairs is concluded from the experimental fact that at the
given background of electrically active defects the Pb-doping do not
appreciably change neither the electron concentration nor the mobility of
the doped films.\cite{ob,obd}

From the thermally activated resistivity observed below room temperature
(Fig.~1) and the large longitudinal magnetoresistivity (Fig.~2), a
percolative mechanism of electrical conduction is gathered. In this
conduction model, electrons are located in potential wells separated by
potential barriers. These potential fluctuations can be related to the
extended structural defects like dislocations, growth flaws, grain
boundaries, etc, as well as to the compensation degree fluctuations related
to the point defects. Within such a model one can expect the macroscopic
percolative electron mobility (the one measured in the Hall experiments) is
smaller than the microscopic one, limited by the ionized defects scattering
only. This expectation is confirmed by the present experiments. Namely, in
order to fit the theoretical curves for magnetoresistivity (MR) obtained in
the frame of the WL theory, to experimental points, we have to assume that
the microscopic mobility is higher (see Table~1).

Doping of InSb films with Pb results in pronounced WL effects: at
temperatures lower than 7~K, the resistance rapidly increases (Fig.~1). The
effect magnitude and temperature range is considerably higher than those
observed previously in pure InSb.\cite{dynes,morita,mani} The relation of
the WL effect to the presence of Pb impurities seems to be further confirmed
by the positive magnetoresistance observed at the lowest magnetic fields
(Fig.~3), which is apparently due to the SO scattering. This effect clearly
demonstrates that the electrons are scattered from heavy atoms. In the
previous investigations only negative MR was observed (no SO interaction has
been found). As far as we know, this is the first observation of SO
scattering of electrons in the conduction band of a 3D semiconductor sample.

However, it is not quite clear in which way the doping with lead could
enhance the WL effect. We suggest that it can be associated with short-range
potentials of the substitutionally incorporated Pb atoms. This problem has
to be clarified in further studies.

To interpret the low temperature MR in our samples, we had to adopt the
theoretical expressions of Kawabata\cite{kawabata} to the particular case of
InSb. We generalized the formulae to take into account both SO coupling and
Zeeman splitting, Eqs.~(2) and (3). From the curve fitting we obtained the
SO dephasing time $\tau _{so}\simeq 10^{-12}$~s. This appears to be a
reasonable value for scatterings on heavy atoms like Pb.

The fitting also supplied the temperature dependence of the phase relaxation
time $\tau _{\varphi }$ shown in Fig.~4. A particular feature of that
dependence is saturation for $T\rightarrow 0$. We show that this peculiar
temperature dependence can be explained if one assumes that $\tau _{\varphi
} $ is a combination of two inelastic relaxation times $\tau _{i}$ and $\tau
_{c}$ (Eq.~(5)). The inelastic relaxation $\tau _{i}$ given by Eq.~(6) is
associated with electron-phonon interaction and as such is an intrinsic
property of InSb. The obtained magnitude and temperature dependence of
$\tau _{\varphi }$ is in a good agreement with that obtained in
Ref.~[\onlinecite{dynes}].

The inelastic relaxation time $\tau _c$ is assumed to be independent on
temperature. In our samples it is in the range $\tau _c=(1$ to $7)\times
10^{-12}$~s. We suppose that it is associated with extended structural
defects of InSb thin films, such as the grain boundaries. In this model the
conductivity electrons are dephased by inelastic scatterings from electrons
occupying the interface states, which have a nonvanishing density of states
at the Fermi level. In such a model $\tau _c$ can be related to the grain
size $L_c$ by $\tau _c\sim L_c^2/D$ ($D$ the diffusion constant). For the
obtained values of $\tau _c$ one obtaines $L_c\simeq 100$~nm. This is a
magnitude of correct order in our samples.

It should be mentioned that our dependence $\tau _{\varphi }(T)$ resembles
that found by Lin and Kao\cite{lin} for a large group of disordered
polycrystalline metals. It is clear that our samples fall into this
category. An interesting finding of Lin and Kao is also that many
disordered metals have the same saturation dephasing length of about 100~nm.
This value coincide with the value of our $L_c$. The saturation observed in
the temperature dependence of the dephasing time has been recently
interpreted in terms of dynamical two-level systems associated with
structural defects such as interfaces, dislocations, etc. These novel
sources of dephasing are thought to be intrinsic to all samples with
structural disorder.
The obtained by us result $\tau _{i}=T^{-3}$ is in disagreement with that of
Ref.~[\onlinecite{mani}] where a dependence of the type $\tau _{i}=T^{-2}$
has been found.

The postulated temperature dependence of $\tau _{\varphi }$ (solid lines in
Fig.~4) was successfully used for the interpretation of the temperature
dependence of the sample resistance above 2~K (Fig.~5). At lower
temperatures, the temperature dependence of the resistance can be explained
by the electron-electron interaction. We wish to draw attention to the fact
that in the localization regime we were able to interpret both the
temperature dependence of the MR and the resistance with the help of the
same temperature dependence of inelastic time $\tau _{\varphi }$ and of the
other parameters. This strengthens the parameter determination because the
two physical effects are different and they are described by different
formulae.

\section*{Acknowledgements}

We thank Dr.~T.~Cichorek and Dr.~A.~Marucha from the Laboratory of Low
temperatures and Structural Investigations PAS in Wroclaw for performing the
measuments below 2~K. One of authors (V.D.) is thankful to the Institute of
Physics of the Pozna\'n Technical University for kind hospitality and to
B.~L.~Altshuler for discussing several points of this work during the
Localization'99 conference. This work was supported by P.P. Grant BW 62-184
and partly by INTAS Grant 2000-0476 and by NATO fellowship grant
CP(UN)06/B/2001/PO.

\end{multicols}

\vskip2cm 

\begin{table}[hbtp]
\begin{tabular}{|c||c|c|c|c|c|c|c|c|}
No & $d, \mu $m & $\mu $, cm$^2$/V$\cdot $s & $n\times 10^{-16}$ cm$^{-3}$ &
$l=v_F\tau $, nm & $1/k_Fl$ & $r_s=1/a_Bk_F$ & $B_c$, T & $l_{\varphi }$, nm
\\ \hline
732 & 2.1 & 1080 (3500) & 1.6 (1.6) & 5.5 (18) & 2.3 (0.72) & 0.19 (0.19) &
21.3 (2.0) & 34 (62) \\ \hline
727 & 2.1 & 760 (830) & 50 (15) & 12 (9) & 0.34 (0.68) & 0.06 (0.09) & 4.4
(8.2) & 90 (64) \\ \hline
725 & 1.6 & 2770 (3000) & 13 (1.9) & 27 (16) & 0.25 (0.75) & 0.10 (0.18) &
0.9 (2.5) & 106 (61)
\end{tabular}
\vskip0.5cm
\caption{Electrical properties of the InSb$<$Pb$>$ films at 4.2~K and
calculated localization theory parameters. In the parentheses are given the
data obtained from the curve fitting.}
\label{tab}
\end{table}

\newpage
\begin{figure}
\psfig{file=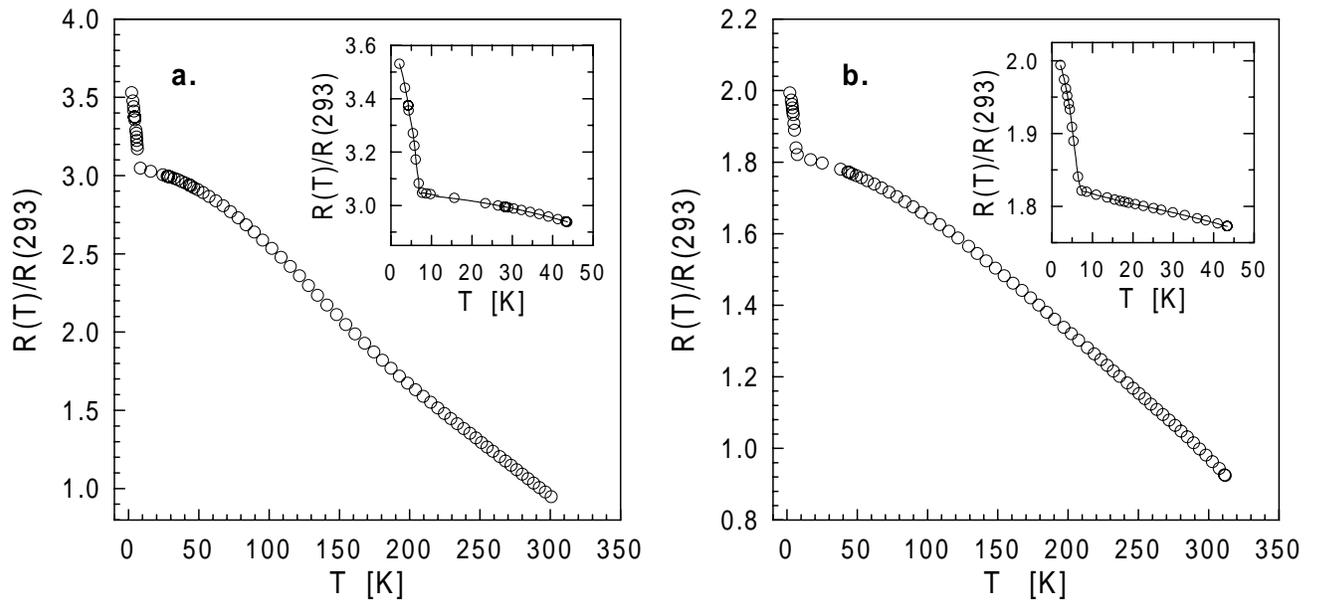,width=20cm}
\caption{The temperature dependence of relative resistivity of InSb$<$Pb$>$ 
films below room temperature. (a) Sample 725B. (b) Sample 727. In the inserts 
the sharp increase in resistivity ascribed to the weak localization effect is
presented in more details.}
\end{figure}

\newpage
\begin{figure}
\psfig{file=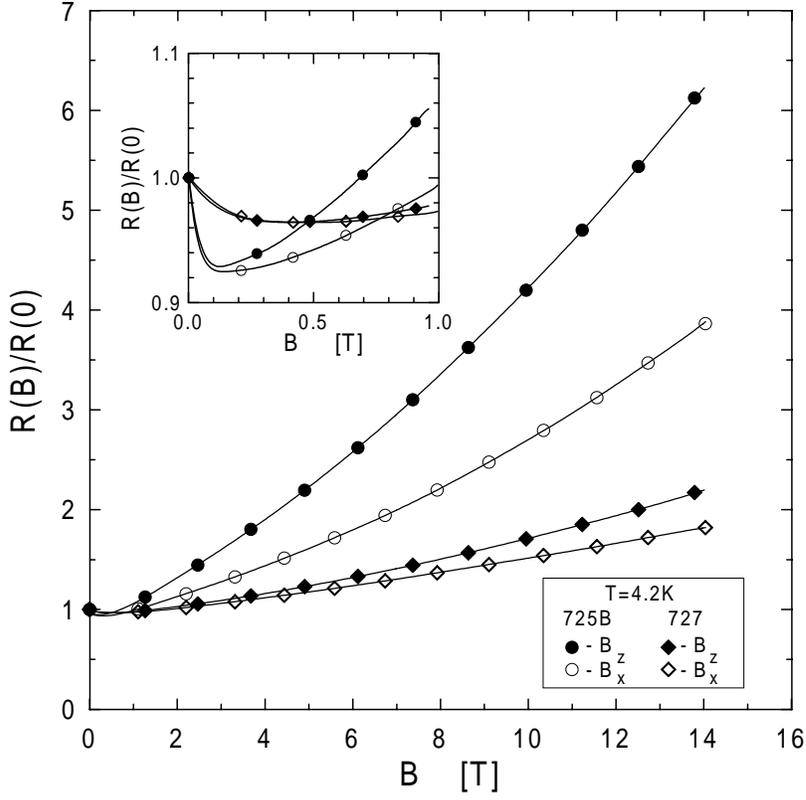,width=14cm}
\caption{The magnetic field
dependence of the magnetoresistivity of samples 725B and 727 at 4.2~K. $B_z$
and $B_x$ are the directions of magnetic field towards the sample. At $B_z$
the transversal MR is measured and at $B_x$ the longitudinal MR is measured.}
\end{figure}

\newpage
\begin{figure}
\psfig{file=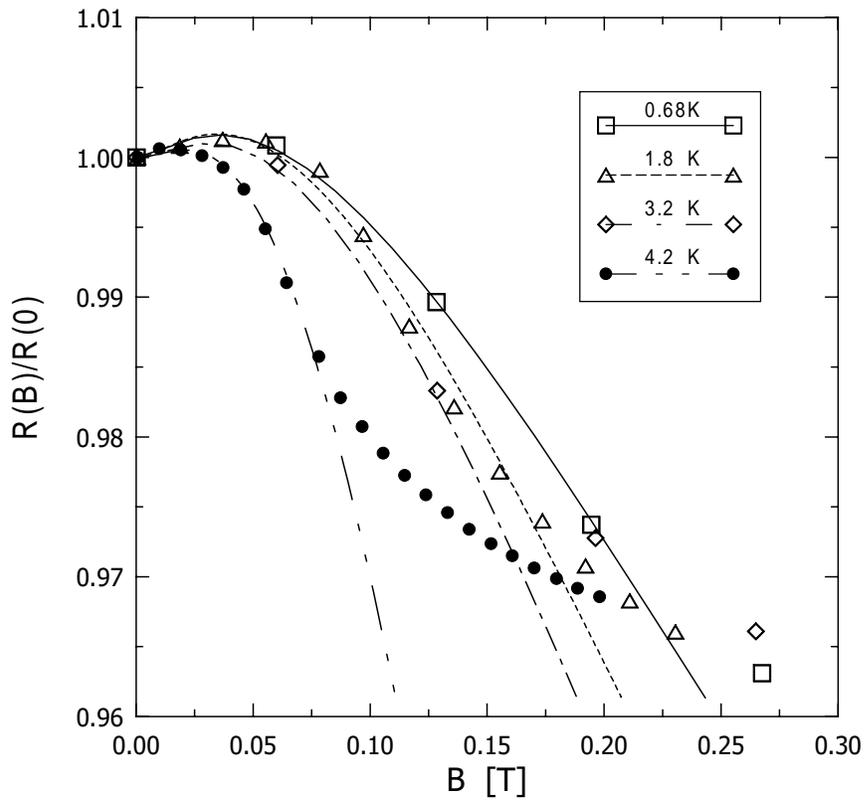,width=14cm}
\caption{Low field magnetoresistivity of sample 727 at
various temperatures. The curves are fittings of the WL theory to the
experimental points.}
\end{figure}

\newpage
\begin{figure}
\psfig{file=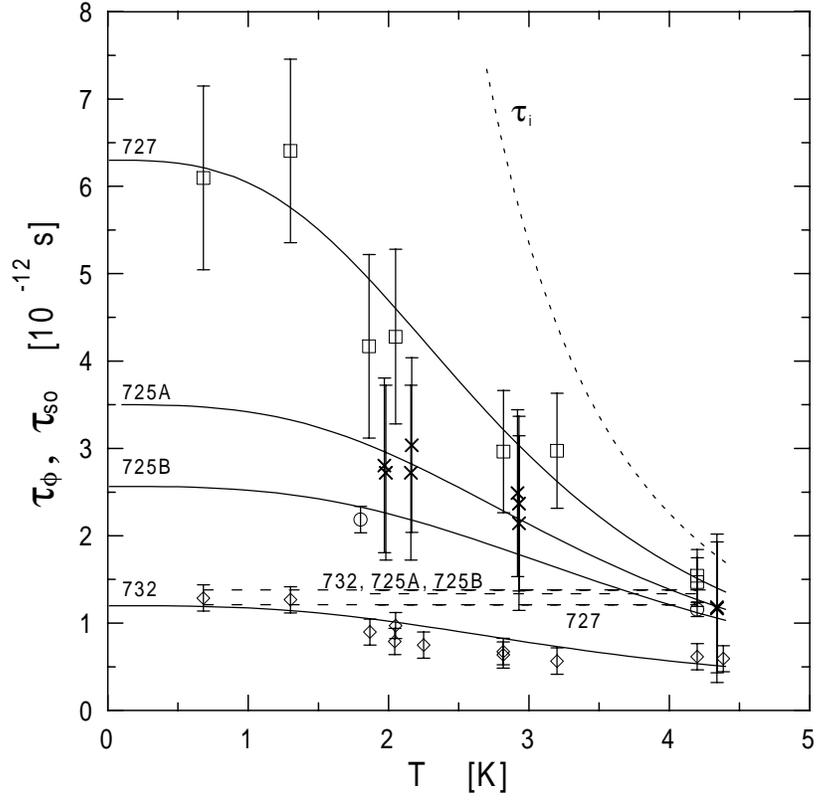,width=14cm}
\caption{Temperature dependencies
of the inelastic relaxation times for samples 725A, 725B, 727 and 732 as
obtained from the theoretical curve fittings to the experimental MR data.
The solid curves are the values of $\tau _{\varphi }$ calculated from
Eq.~(5), the dashed lines are the values of $\tau _{so}$, and the dotted
line describes $\tau _i$ calculated from Eq.~(6).}
\end{figure}

\newpage
\begin{figure}
\psfig{file=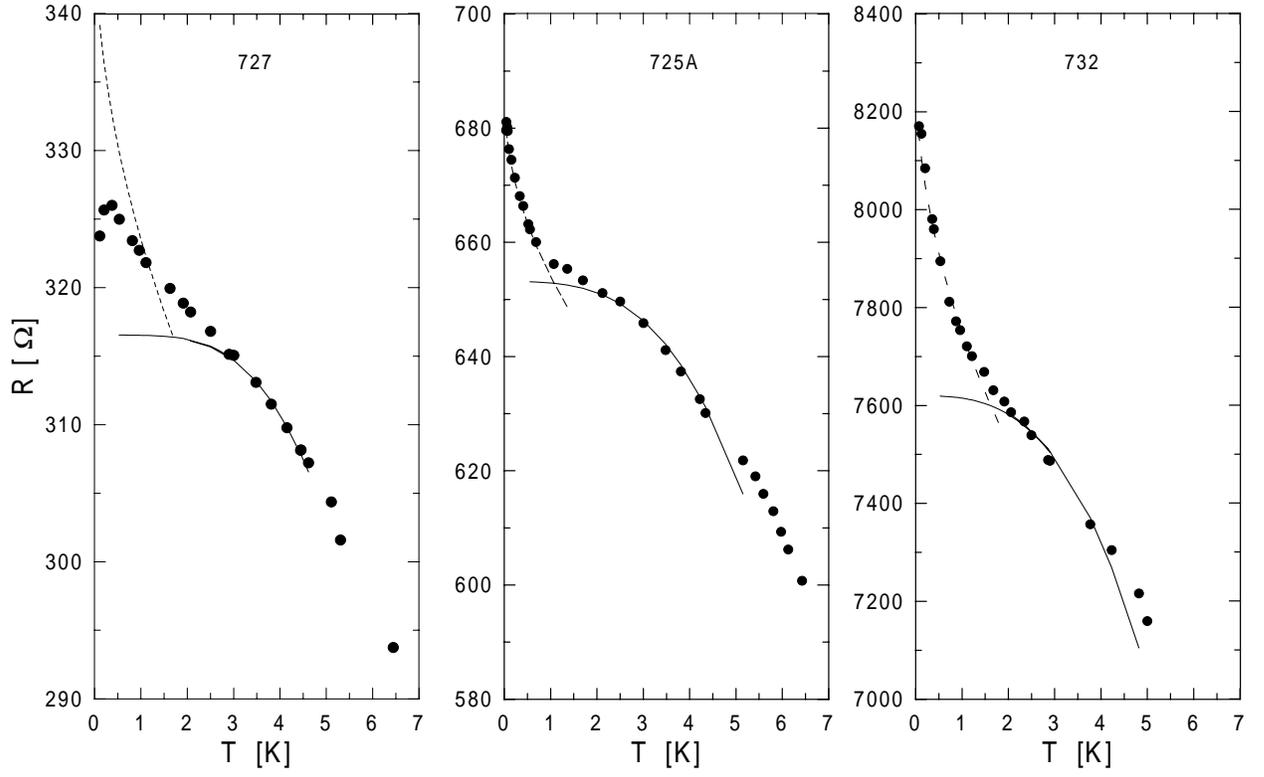,width=19cm}
\caption{The temperature dependence of resistance of samples 727, 725 and 732
between 40~mK and 7~K. The solid lines are calculated from Eqs.~(8) and (9)
as explained in the text.}
\end{figure}

\end{document}